\documentclass[letter,twocolumn]{jpsj2}
\def\runtitle{Y. Hasegawa and M. Yakiyama}
\def\runauthor{Horizontal Line Nodes in Sr$_2$RuO$_4$}
\title{
Interband Coulomb Interaction and
Horizontal Line Nodes in Triplet Superconductor
 Sr$_2$RuO$_4$
}

\author{Yasumasa Hasegawa
 and Mayumi Yakiyama}
\inst{
Department of Material Science,
Graduate School of Science,\\
Himeji Institute of Technology
 Ako, Hyogo 678-1297, Japan}

\recdate{\today}

\abst{
A possible mechanism for the horizontal line nodes in
triplet superconductor, Sr$_2$RuO$_4$, is proposed.
 We consider the interlayer Coulomb interaction,
as well as the on-site Coulomb
repulsion, between
electrons in different bands.
In the second order perturbation
 in the interband interaction,
the effective interaction becomes dependent on $\cos \frac{q_z}{2}$,
resulting in the horizontal line nodes.
}

\kword{%
triplet superconductivity, line nodes, Hubbard model, 
Sr$_2$RuO$_4$}

\begin{document}
\maketitle



Since the discovery of the superconductivity in Sr$_2$RuO$_4$
by Maeno et al.\cite{Maeno94},
its unique properties have been revealed by many experiments.
Experiments on Knight shift\cite{Ishida98}
 and elastic neutron scattering\cite{Duffy2000} show that 
the electrons make the triplet pairs with the $\mib{d}$-vector
parallel to $\hat{z}$ as predicted 
by Rice and Sigrist\cite{RiceSigrist95}
in analogy with superfluid $^3$He.
The superconductivity breaks the time-reversal symmetry, as indicated by
$\mu$SR experiment\cite{Luke98}.
The energy gap has line nodes as shown by the temperature dependence of 
specific heat\cite{Nishizaki2000}, and the
relaxation rate in  nuclear magnetic resonance (NMR)\cite{Ishida2000}.
The existence of the line nodes is confirmed by other 
experiments\cite{Lupien2001,Suzuki2002,Tanatar2001,Izawa2001}.
The absence of the angle dependence of the 
thermal conductivity in the magnetic field
 within the $a$-$b$ plane \cite{Tanatar2001,Izawa2001}
 shows that the line nodes run horizontally on the
Fermi surface.

When we assume that the order parameter depends only on $k_x$ and $k_y$,
the phenomenological theory based on the group theory in the  
point group of D$_{4h}$ shows that existence of line nodes and 
the breakdown of the time-reversal symmetry  are in general 
incompatible with the symmetry and that only accidental
vertical line nodes are possible.\cite{Hasegawa2000}
Quasi-two-dimensional nature of the Fermi surface, however,
should be taken into account.\cite{Hasegawa2000}
 The Fermi surface in Sr$_2$RuO$_4$ consists of
three cylindrical sheets, which are open in the $k_z$ direction.
In this case the order parameter should be expanded in the Fourier series 
rather than Taylor series (s-, p-, d- partial waves) in $k_z$.
In the presence of the interlayer interaction, the horizontal line nodes 
are possible.\cite{Hasegawa2000}

Triplet superconductivity is shown to be 
caused  when spin fluctuation is anisotropic and the 
Fermi surface is one-dimensional\cite{Sato2000,Kuwabara2000}.
Takimoto\cite{Takimoto2000} showed that triplet superconductivity appears
in the three-band Hubbard model
when the on-site Coulomb interaction between electrons in different bands
is large.
Nomura and Yamada\cite{Nomura2002}
 studied the three-band Hubbard model in the perturbation theory
up to third order. By solving the linearized Eliashberg equation
numerically, they obtained that the triplet-superconductivity
with the vertical line-node-like structure  in
the $\beta$ band  is 
stabilized.
The vertical line nodes, however, can be wiped out by the mixing of the 
order parameters compatible with the symmetry 
at $T<T_c$.\cite{Hasegawa2000,Zhitomirsky2001}
Since they consider only two-dimensional 
model\cite{Sato2000,Kuwabara2000,Takimoto2000,Nomura2002},
the possibility of the horizontal line node has not been studied. 

The superconductivity with horizontal line nodes are
studied by
assuming the interlayer attractive 
interaction\cite{Hasegawa2000,Kubo2000,Kuboki2001,Annett2002}.
Zhitomirsky and Rice\cite{Zhitomirsky2001}
 has proposed the pair hopping model between bands, 
which they call
interband proximity effect. 
In that model the active band has full gap on the Fermi surface, while
the passive band has line nodes.
They argued that there exists
 the pair hopping term from the pair at $\mib{r}$ and
$\mib{r} + (a,0,0)$ in the active band to the pair at $\mib{r}$ and
$\mib{r} + (a/2,a/2,c/2)$ in the passive band,
where $a$ and $c$ are the lattice constants.
The position of line nodes has not been studied experimentally yet.
We have proposed that the existence of the horizontal line nodes in the
nested part of the Fermi surface can be observed 
by the inelastic neutron experiment\cite{Yakiyama2003}.

Recently,
Kondo\cite{Kondo2001}
 studied the two-dimensional Hubbard model at $T=0$ in the second 
order perturbation theory. He obtained that singlet superconductivity is
stabilized in the wide range of electron filling.

In this paper we
propose a mechanism for the horizontal line node in Sr$_2$RuO$_4$
by applying the Kondo's approach to the multiband extended Hubbard model.
We show that the effective interaction resulting in  the line nodes
can be derived from  the lowest order in the 
interlayer interband Coulomb interaction.

The  
 interband Coulomb interaction is written as
\begin{align}
 {\cal H}_{\textrm{IB}}
&= 
\sum_{\mib{k},\mib{k}',\mib{k}''}
\sum_{\sigma,\sigma'}\sum_{l\neq l'}
V_{\textrm{IB}}(\mib{k}-\mib{k}')
\nonumber \\
&\times
 c^{\dagger}_{\mib{k}',l, \sigma} c_{\mib{k},l ,\sigma }
 c^{\dagger}_{\mib{k}''+\mib{k}-\mib{k}',l',\sigma'}c_{\mib{k}'',l',\sigma'} ,
\end{align} 
where $l$ is the band index
and
\begin{align}
V_{\textrm{IB}}(\mib{k}-\mib{k}')&=
V \left(1+\alpha
\cos \frac{a (k_x-k_x')}{2}
\cos \frac{a (k_y-k_y')}{2}
\right.
\nonumber \\
&\times \left.
\cos \frac{c (k_z-k_z')}{2}\right) .
\end{align}
In the above $V$ is the on-site interband interaction and $\alpha V$
is the interlayer interband interaction.

\begin{figure}[tb]
\includegraphics[width=0.45\textwidth]{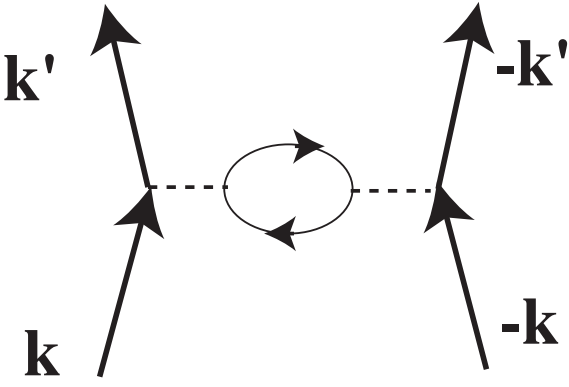}
 \caption{Effective interaction in the second order in
the interband interaction 
$V_{\textrm{IB}}$.
Thick and thin lines are the Green's functions for the band $l$ and $l' \neq l$,
respectively. Dotted lines are $V_{\textrm{IB}}$.
}
\label{fig1}
\end{figure}

The effective interaction in the second order in $V_{\textrm{IB}}$ 
is shown in 
Fig.~\ref{fig1}.
Note that  singlet superconductivity and triplet superconductivity have the 
same form in this model.
We get
\begin{align}
V_{\textrm{eff},l}(\mib{q})&=
-\chi_{l'}(q_x, q_y)
\left( V_{IB}(\mib{q})\right)^2 + U
\nonumber \\
&=
V_0(q_x,q_y)
+2V_1(q_x,q_y) \cos\frac{cq_z}{2}
\nonumber \\
&
+2V_2(q_x,q_y) \cos cq_z ,
\label{eq3}
\end{align}
where 
$\mib{q}=\mib{k}-\mib{k}'$, $U$ is the on-site intraband 
Coulomb interaction,
\begin{equation}
V_0(q_x, q_y)=-\chi_{l'}(q_x,q_y) (1+\frac{\alpha^2}{2}
\cos^2\frac{aq_x}{2} \cos^2\frac{aq_y}{2})+U ,
\end{equation}
\begin{equation}
V_1(q_x,q_y)=-\alpha \chi_{l'}(q_x,q_y) \cos\frac{a q_x}{2} 
 \cos \frac{a q_y}{2} ,
\end{equation}
\begin{equation}
V_2(q_x,q_y)=-\frac{\alpha^2}{4}\chi_{l'}(q_x,q_y)\cos^2\frac{aq_x}{2}
\cos^2\frac{aq_y}{2}
\end{equation}
 and
$\chi_{l'}(\mib{q})$ is the susceptibility in the $l'$ band.
We have neglected the $k_z$ dependence in $\epsilon_{\mib{k} l'}$, 
since the Fermi surface has little warping  in Sr$_2$RuO$_4$.
Then  $\chi_{l'}(\mib{q})$  is independent of $q_z$ and the 
$q_z$ dependence of $V_{\textrm{eff},l}(\mib{q})$ comes 
from the interlayer interaction.
We have also neglected the second order 
perturbation in $U$ and 
 the Coulomb interaction between electrons
at the nearest sites in the same plane. 
The interaction within the same plane changes only $V_0(k_x,k_y)$ and
does not affect $V_1(k_x,k_y)$ and $V_2(k_x,k_y)$  in
Eq.(\ref{eq3}). Since we are interested in the
mechanism for the horizontal line nodes,
 we have neglected these terms.

We write the two-dimensional intersection of the
Fermi surface as a function of $\theta$, i.e.
$\mib{k}$ on the Fermi surface is written as 
$(k_F(\theta) \cos \theta, k_F(\theta) \sin \theta, k_z)$.
As shown by Kondo\cite{Kondo2001}, 
the most stable state at $T=0$ is given by the solution
\begin{equation}
-\int_0^{2\pi} d\theta' {V}_{Fi}(\theta, \theta') 
\rho(\theta') z(\theta')=\lambda z(\theta)
 \hspace{0.5cm} \mbox{($i=0,1,2$)}
\label{eigenvalue1}
\end{equation} 
with the largest eigenvalue $\lambda$,
where  
\begin{align}
{V}_{Fi}(\theta, \theta')=V_i(
&k_F(\theta)\cos\theta-k_F(\theta')\cos\theta', \nonumber \\
& k_F(\theta)\sin\theta-k_F(\theta')\sin\theta') ,
\end{align}
$ \rho(\theta')$ is the density of states on the Fermi surface and
$z(\theta)$ is the momentum dependence of the
 order parameter on the Fermi surface ($\Delta(\mib{k})=\Delta_0 z(\theta)$,
$\Delta_0 z(\theta) \cos\frac{ck_z}{2}$, or 
$\Delta_0 z(\theta) \cos {ck_z}$
for the singlet superconductivity and $d_z(\mib{k})=\Delta_0 z(\theta)$,
$\Delta_0 z(\theta) \cos\frac{ck_z}{2}$, or
$\Delta_0 z(\theta) \cos {ck_z}$
for the triplet superconductivity with $\mib{d}(\mib{k}) 
\parallel \hat{\mib{z}}$).
By defining
\begin{equation}
w(\theta)=\sqrt{\rho(\theta)}z(\theta)
\end{equation}
Eq.~(\ref{eigenvalue1}) becomes symmetric form as
\begin{equation}
-\int_0^{2\pi} d\theta' \tilde{V}_{Fi}(\theta, \theta') 
 w(\theta')=\lambda w(\theta)
 \hspace{0.5cm} \mbox{($i=0,1,2$)}
\label{eigenvalue2}
\end{equation} 
where \begin{equation}
\tilde{V}_{Fi}(\theta,\theta')=
{V}_{Fi}(\theta,\theta')\sqrt{\rho(\theta)\rho(\theta')}
\end{equation}

The similar equation 
is obtained by maximizing the average of the effective interaction
on the Fermi surface with respect to the order parameter
\begin{equation}
\frac{\delta \langle \tilde{V}_{Fi}(\theta,\theta')\rangle }
{\delta w(\theta)} = 0
\label{variation}
\end{equation}
where
\begin{equation}
\langle \tilde{V}_{Fi}(\theta, \theta') \rangle
= \frac{\int\!\!\int d\theta d\theta'
 \tilde{V}_{Fi}(\theta,\theta') 
w(\theta)w(\theta')}
{\int d\theta   (w(\theta))^2}
\end{equation}
is the average of the effective interaction for the order parameter
$z(\theta)$ on the Fermi surface.
Eq.~(\ref{variation}) is written as
\begin{equation}
\int_0^{2\pi} \tilde{V}_{Fi}(\theta,\theta') w(\theta') d\theta'
= \langle \tilde{V}_{Fi}(\theta,\theta') \rangle w(\theta)
\label{eigenvalue3}
\end{equation}
Comparing Eq.(\ref{eigenvalue3}) with Eq.~(\ref{eigenvalue2}),
we get
\begin{equation}
\lambda=- \langle \tilde{V}_{Fi}(\theta,\theta') \rangle
\end{equation}

Using the Fourier expansion for $z(\theta)$ we get 
 the eigenvalue problem
of the non-Hermite matrix\cite{Kondo2001} from Eq.~(\ref{eigenvalue1}).
We can obtain the eigenvalue problem with Hermite matrix
from  Eq.~(\ref{eigenvalue2}),
\begin{equation}
 \sum_{l'} M_{ll'}^{(i)} w_{l'}=\lambda w_{l}, 
\end{equation}
where 
\begin{equation}
M_{ll'}^{(i)}=-\frac{1}{2\pi}\int_0^{2\pi}\!\!\int_0^{2\pi}d\theta d\theta'
\tilde{V}_{Fi}(\theta,\theta') e^{i(l'\theta'-l\theta)} ,
\end{equation}
and
\begin{equation}
w(\theta)=
\sum_{l}w_l e^{i l \theta} .
\end{equation}

In Sr$_2$RuO$_4$, 
the $\beta$ and the $\gamma$ band have the cylindrical Fermi surfaces with
the cross section of 0.457 and 0.667 of the Brillouin zone, 
respectively\cite{Mackenzie96}.
As a first step, we  use the  simple model that two bands have the
same energy dispersion.
We take the simple model that there are two bands which have the same energy 
\begin{align}
& \epsilon_l(\mib{k})=
\epsilon_{l'}(\mib{k}) \nonumber \\
=&
-2t \left( \cos a k_x + \cos ak_y\right)
 -4t' \cos ak_x \cos a k_y -\mu
\end{align}
with $t=1$ and $t'=0.32$.

 The representation of the D$_{4h}$ point group is given by
four one-dimensional representations, $a_1$, $a_2$, 
$b_1$, $b_2$ and one two-dimensional
representation $e$.
In each one-dimensional representation the eigenvector satisfies the relation,
\begin{align}
w(\theta)&=\eta_{2}w(\theta+\pi)=\eta_{4}w(\theta+\frac{\pi}{2})\nonumber \\
&=\eta_{U} w(-\theta) =\eta_{U'} w(-\theta+\frac{\pi}{2})
\end{align}
where $\eta_{2}$, $\eta_{4}$, $\eta_{U}$ and $\eta_{U'}$
 are the characters of the $\pi$-rotation around the $z$-axis,
the $\frac{\pi}{2}$-rotation around the $z$-axis,
 the $\pi$-rotation around the $x$-axis,
and the $\pi$-rotation around the $x=y$ line, respectively
for 
each one-dimensional irreducible representations.
The irreducible representations $a_1$, $a_2$, $b_1$, and $b_2$ behave as 
extended $s$-wave, $g$-wave ($k_x k_y (k_x^2-k_y^2)$),
$k_x^2-k_y^2$, and $k_x k_y$, respectively. 

We exclude the constant term $w_{l=0}$,
 which should be included in the $a_1$ symmetry, 
in the $k_z$-independent order parameter 
in order to avoid  a large intraband on-site interaction $U$.
Since $\rho(\theta)$ is not constant, the effect of $U$
is not completely avoided by removing $w_0$.
The   intraband on-site interaction, however,
changes  only superconductivity with the $k_z$-independent
$a_1$ symmetry which is not very important in this study
as we show below. Thus we use the 
approximation of removing $w_0$.

For the two-dimensional representation, 
$w(\theta)$ should satisfy
\begin{equation}
w(\theta)=-w(\theta+\pi)=\pm w(-\theta).
\label{2drep}
\end{equation}
In the present approximation that only
the lowest order in the order parameter
is considered, two-dimensional representation corresponds to two 
degenerate states for $\pm$ in Eq. (\ref{2drep}),
which behaves as  $k_x$ and $k_y$, respectively.
For example, the $e$ symmetry for $\tilde{V}_{F1}(\theta,\theta')$
has the  the order parameter represented by
\begin{equation}
d_z(\mib{k}) \approx \Delta_0  k_x \cos\frac{ck_z}{2},  
\Delta_0  k_y \cos\frac{ck_z}{2}
\end{equation}
or
\begin{equation}
\Delta(\mib{k}) \approx \Delta_0  k_x \sin\frac{ck_z}{2},
\Delta_0  k_y \sin\frac{ck_z}{2}
\end{equation}
In our approximation triplet and singlet superconductivity have the same
stability. 
The actual order parameter in the two-dimensional representation
 is expected to be
\begin{equation}
d_z(\mib{k}) \approx \Delta_0  \left( k_x + i k_y \right ) \cos\frac{ck_z}{2}
\end{equation}
or
\begin{equation}
\Delta(\mib{k}) \approx \Delta_0 \left( k_x + i k_y \right) \sin\frac{ck_z}{2},
\end{equation}
if the higher order terms in the order parameter
 is taken into account for the superconducting condensation energy.

\begin{figure}[tb]
\includegraphics[width=0.45\textwidth]{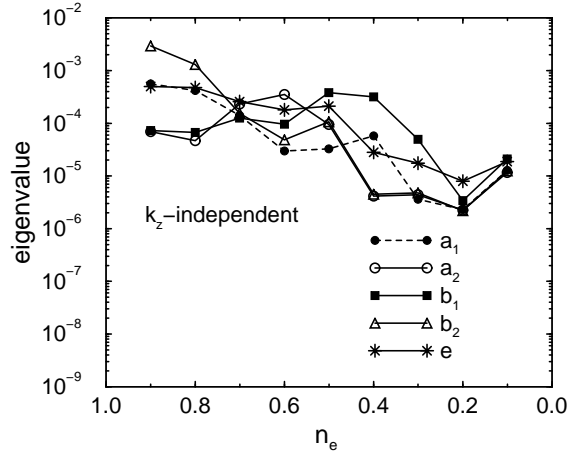}
 \caption{
Maximum eigenvalues of $\tilde{V}_{F0}(\theta,\theta')$
 as a function of electron density $n_e$ for $\alpha=0.1$.}
\label{fig2}
\end{figure}
\begin{figure}[tb]
\includegraphics[width=0.45\textwidth]{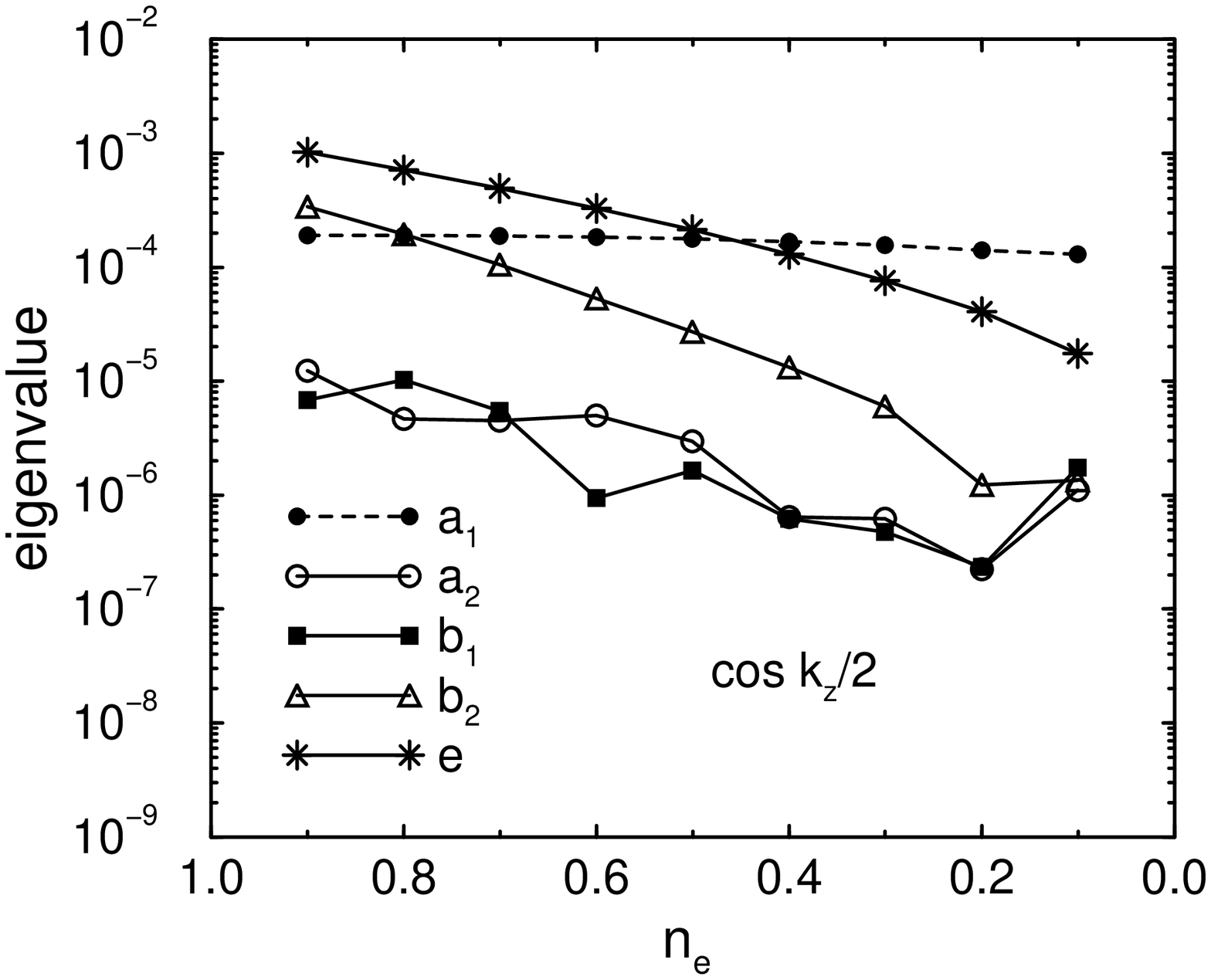}
 \caption{
Maximum eigenvalues of $\tilde{V}_{F1}(\theta,\theta')$
 as a function of electron density $n_e$ for $\alpha=0.1$.
}
\label{fig3}
\end{figure}
\begin{figure}[tb]
\includegraphics[width=0.45\textwidth]{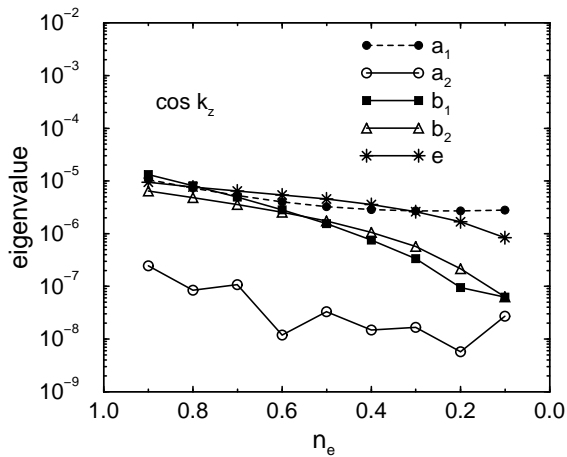}
 \caption{
Maximum eigenvalues of $\tilde{V}_{F2}(\theta,\theta')$
 as a function of electron density $n_e$ for $\alpha=0.1$.
}
\label{fig4}
\end{figure}
In Figs.~\ref{fig2}-\ref{fig4}
we plot the maximum eigenvalues for each irreducible representation
 as a function of the electron density $n_e$
for $\alpha=0.1$.
It is seen from Figs.~\ref{fig2}-\ref{fig4} that 
the superconductivity with
$e\times\cos\frac{ck_z}{2}$ symmetry (and $e\times\sin\frac{ck_z}{2}$ 
symmetry, which  degenerates with $e\times\cos\frac{ck_z}{2}$ 
in the present model) has the maximum eigenvalue for 
$0.8 \gtrsim n_e \gtrsim 0.5$.

\begin{figure}[tb]
\includegraphics[width=0.45\textwidth]{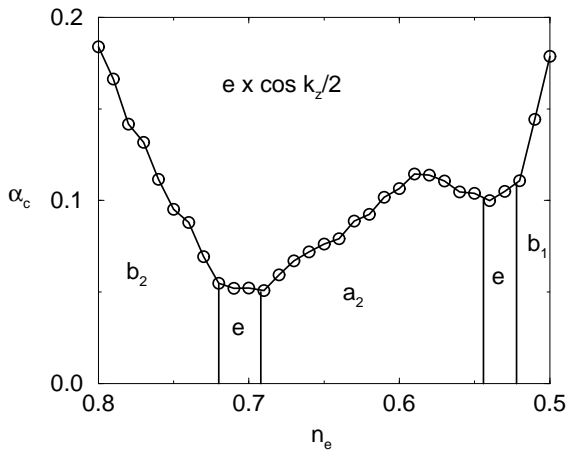}
 \caption{
Phase diagram for the extended interband Hubbard model with 
$t=1$ and $t'=0.32$.
}
\label{fig5}
\end{figure}

In Fig.~\ref{fig5} we show the phase diagram in the $n_e$-$\alpha$ plane.
The triplet superconductivity with horizontal line nodes 
(e $\times \cos \frac{ck_z}{2}$) is 
 stabilized 
in some region of parameters, which may be realized in Sr$_2$RuO$_4$.

We have shown that the 
effective interaction due to the interband interaction can cause the 
superconductivity with horizontal line nodes.
The mechanism for the superconductivity
proposed in the present paper
is  the extension of the idea by Little\cite{Little1964},
who proposed the  attractive interaction between electrons
via a side-chain Coulomb interaction. In our case the different band plays
the role of the side-chain.

The degeneracy between $e\times\cos\frac{ck_z}{2}$ and $e\times\sin\frac{ck_z}{2}$
can be lifted by several effects such as warping of the Fermi surface
due to interlayer hopping\cite{Kuboki2001}, 
intra-band interactions, 
higher order perturbation, anisotropy of the spin susceptibility, etc.

The order parameter with horizontal line-node is stabilized by the relatively small 
interlayer interaction ($\alpha \approx 0.1$).
The anisotropic superconductivity due to the second order
perturbation in the on-site interaction is 
caused by the variation of $\chi(\mib{q})$ from its average
in order to avoid the large on-site Coulomb repulsion $U$.\cite{Kondo2001}
 On the other hand,   the product of the on-site interaction 
and the interlayer interaction 
can use the full value of the susceptibility
to get the superconductivity. This is why 
the superconductivity with horizontal line nodes is stabilized by 
a small 
interlayer interaction.

We have shown that 
the triplet superconductivity with horizontal line nodes, i.e. 
  $e \times \cos \frac{k_z}{2}$ state, is shown to be stabilized
in a reasonable parameter range.
We have neglected the fact that
$\chi(\mib{q})$ is large due to the nesting nature of the
$\alpha$ and $\beta$ bands\cite{Mazin99,Sidis99}.
If we consider the enhancement of $\chi(\mib{q})$, 
the effective interaction in the $\gamma$ band becomes large 
due to the mechanism discussed in this paper.

The interband proximity effect\cite{Zhitomirsky2001}
is compatible with the effective interaction studied in this paper.
When both interband interlayer interaction and
interband proximity effect
are considered,
the superconductivity with the horizontal line nodes
will be much favored.

%


\begin{thebibliography}{99}

\bibitem{Maeno94}
Y. Maeno, H. Hashimoto,K. Yoshida, S. Nishizaki, T. Fujita,J.G. Bednorz, and
F. Lichitenberg,
Nature (London) \textbf{372} (1994) 532.

\bibitem{Ishida98}
 K. Ishida, H. Mukuda, Y. Kitaoka, K. Asayama, Z. Q. Mao, Y. Mori and Y. Maeno,
 Nature (London) \textbf{396} (1998) 658.

%
\bibitem{Duffy2000}
J. A. Duffy, S. M. Hayden, Y. Maeno, Z. Mao, J. Kulda, and G. J. McIntyre,
 Phys. Rev. Lett. \textbf{85} (2000) 5412.
%
\bibitem{RiceSigrist95}
T.M. Rice and M. Sigrist, J. Phys. Condens. Matter, \textbf{7} (1995) L643.


\bibitem{Luke98}
G. M. Luke, Y. Fudamoto, K. M. Kojima, M. I. Larkin, J. Merrin, B. Nachumi, 
Y. J. Uemura, Y. Maeno, Z. Q. Mao, Y. Mori, H. Nakamura, and M. Sigrist,
 Nature (London) \textbf{394} (1998) 558.


\bibitem{Nishizaki2000}
S. NishiZaki, Y. Maeno and Z. Q. Mao,
J. Phys. Soc. Jpn. \textbf{69} (2000) 572.


\bibitem{Ishida2000}
K. Ishida, H. Mukuda, Y. Kitaoka, Z. Q. Mao, Y. Mori, and Y. Maeno
Phys. Rev. Lett. \textbf{84} (2000) 5387.


\bibitem{Lupien2001}
C. Lupien, W. A. MacFarlane, Cyril Proust, and Louis Taillefer
Phys. Rev. Lett. \textbf{86} (2001) 5986.


\bibitem{Suzuki2002}
M. Suzuki, M. A. Tanatar, N. Kikugawa, Z. Q. Mao, Y. Maeno, and T. Ishiguro,
Phys. Rev. Lett. \textbf{88} (2002) 227004.



\bibitem{Tanatar2001}
M. A. Tanatar, M. Suzuki, S. Nagai, Z. Q. Mao, Y. Maeno, and T. Ishiguro,
Phys. Rev. Lett. \textbf{86} (2001) 2649.

\bibitem{Izawa2001}
K. Izawa, H. Takahashi, H. Yamaguchi, Y. Matsuda, M. Suzuki, T. Sasaki,
 T. Fukase, Y. Yoshida, R. Settai, and Y. Onuki,
 Phys. Rev. Lett. \textbf{86} (2001) 2653.


\bibitem{Hasegawa2000}
Y. Hasegawa, K. Machida, and M. Ozaki, J. Phys. Soc. Jpn. \textbf{69},
(2000) 336.

\bibitem{Sato2000}
M. Sato and M. Kohmoto, J. Phys. Soc. Jpn. \textbf{69} (2000) 3505.


\bibitem{Kuwabara2000} 
T. Kuwabara and M. Ogata, Phys. Rev. Lett. \textbf{85} (2000) 4586.


\bibitem{Takimoto2000}
T. Takimoto, Phys. Rev. B \textbf{62} (2000) R14641.


\bibitem{Nomura2002}
T. Nomura and K. Yamada, J. Phys. Soc. Jpn. \textbf{71} (2002) 404.


\bibitem{Zhitomirsky2001}
M. E. Zhitomirsky and T. M. Rice, Phys. Rev. Lett. \textbf{87} (2001) 057001.



\bibitem{Kubo2000}
K. Kubo and D.S. Hirashima, J. Phys. Soc. Jpn. \textbf{69} (2000) 3489.

\bibitem{Kuboki2001}
K. Kuboki, J. Phys. Soc. Jpn. \textbf{70} (2001) 2698.



\bibitem{Annett2002}
J.F. Annett, G. Litak, B.L. Gyorffy and K.I. Wysokinski,
Phys. Rev. B \textbf{66} (2002) 134514.



\bibitem{Yakiyama2003}
M. Yakiyama and Y. Hasegawa,
Phys. Rev. B \textbf{67} (2003) 014512.

\bibitem{Kondo2001}
J. Kondo, J. Phys. Soc. Jpn. \textbf{70}, (2001) 808.








\bibitem{Mackenzie96}
A. P. Mackenzie, S. R. Julian, A. J. Diver, G. J. McMullan, M. P. Ray,
 G. G. Lonzarich, Y. Maeno, S. Nishizaki, and T. Fujita,
Phys. Rev. Lett. \textbf{76}, 3786 (1996).


\bibitem{Little1964}
W. A. Little, Phys. Rev. \textbf{134} (1964) A1417.



\bibitem{Mazin99}
I. I. Mazin and D. J. Singh, Phys. Rev. Lett. \textbf{82} (1999) 4324.

\bibitem{Sidis99}
Y. Sidis, M. Braden, P. Bourges, B. Hennion, S. NishiZaki, Y. Maeno, and Y. Mori,
Phys. Rev. Lett. \textbf{83} (1999) 3320.
















































\end{thebibliography}
\end{document}